\documentclass[aps,pra,amssymb,amsmath,eqsecnum,nofootinbib]{revtex4}
\usepackage{graphicx}
\newtheorem{principle}{Principle}[section]

\newtheorem{definition}{Definition}[section]
\newtheorem{theorem}{Theorem}[section]

\newtheorem{remark}{Remark}[section]

\newenvironment{proof}{\begin{center}PROOF: \end{center}} {$ \blacksquare $}
\newtheorem{example}{Example}[section]
\begin{document}
\title{A Gr\"{o}enewald Van Hove like formulation of the ordering problems of General Relativity}
\author{Gavriel Segre}
\email{Gavriel.Segre@msi.vxu.se}
 \affiliation{International Center
for Mathematical Modelling in Physics and Cognitive Sciences,
University of V\"{a}xj\"{o}, S-35195, Sweden}
\begin{abstract}
A simple formal recasting of well known arguments concerning the
ordering problems of General Relativity allows to obtain in such a
context a Gr\"{o}enewald Van Hove like theorem.
\end{abstract}
\maketitle
\newpage
\section{Introduction}
Ordering problems occurs in two physically completely different
contexts:
\begin{itemize}
    \item in Quantum Mechanics  \cite{Abraham-Marsden-78} one has that in the canonical quantization of
a monomial in q and p of the form $ q^{n} p^{m} \, n, m \in
\mathbb{N}_{+} $ since the classical observables q and p commutes
while $ [ \hat{q} , \hat{p} ] \; = \; i  $, to each of the
classically  equivalent orderings:
\begin{equation}
    q^{n} p^{m} \; = \; q^{n-1} p^{m} q \; = \cdots \; = \; q p^{m}
    q^{n-1} \; = \; p^{m} q^{n} \; = p q^{n} p^{m-1} \; = \;
    \cdots
\end{equation}
corresponds a different quantum operator:
\begin{equation}
   \hat{q}^{n} \hat{p}^{m} \; \neq \; \hat{q}^{n-1} \hat{p}^{m} \hat{q} \; \neq \; \cdots \; \neq \; \hat{q} \hat{p}^{m}
    \hat{q}^{n-1} \; \neq \; \hat{p}^{m} \hat{q}^{n} \; \neq \; \hat{p} \hat{q}^{n}
    \hat{p}^{m-1} \; \neq \; \cdots
\end{equation}
    \item in General Relativity the application of the
     the \emph{Minimum Substitution Prescription}, following the terminology of \cite{Wald-84}, allows to generalize \emph{a non-gravitational law of
Physics} holding on Minkowski spacetime $ ( \mathbb{R}^{4} ,
\eta_{a b}) $ (where in this paper the \emph{abstract index
notation} is adopted \cite{Wald-84}) and containing a tensorial
quantity $ \nabla^{(\eta)}_{a} \nabla^{(\eta)}_{b} X^{a_{1} \cdots
a_{r}}_{b_{1} \cdots b_{s} } $ (where $ X^{a_{1} \cdots
a_{r}}_{b_{1} \cdots b_{s} } \in \mathcal{T}^{r}_{s} (
\mathbb{R}^{4} ) : (r,s) \neq (0,0)$) to an arbitrary curved
space-time $ ( M , g_{ab} ) $ through the ansatz :
\begin{equation}
  (\eta_{a b} \rightarrow g_{a b} \, , \, \nabla^{(\eta)}_{a}
    \rightarrow  \nabla^{(g)}_{a})
\end{equation}
Ordering problems then occur owing to the fact that while
covariant derivatives commute on Minkowski spacetime:
\begin{equation}
  \nabla^{(\eta)}_{a} \nabla^{(\eta)}_{b}  X^{a_{1} \cdots a_{r}}_{b_{1} \cdots b_{s}}  \; = \;   \nabla^{(\eta)}_{b}
  \nabla^{(\eta)}_{a}  X^{a_{1} \cdots a_{r}}_{b_{1} \cdots b_{s}}  \;\; \forall  X^{a_{1} \cdots a_{r}}_{b_{1} \cdots b_{s}}  \in \mathcal{T}^{r}_{s}(
  \mathbb{R}^{4} ) \, , \, \forall  r,s \in \mathbb{N}
\end{equation}
covariant derivatives on a curved spacetime applied to something
different from a scalar field don't commute:
\begin{equation}
    \nabla^{(g)}_{a} \nabla^{(g)}_{b} X^{a_{1} \cdots a_{r}}_{b_{1} \cdots b_{s} } \; \neq \;
\nabla^{(g)}_{b}
  \nabla^{(g)}_{a}  X^{a_{1} \cdots a_{r}}_{b_{1} \cdots b_{s}}  \;\; \forall  X^{a_{1} \cdots a_{r}}_{b_{1} \cdots b_{s}}  \in \mathcal{T}^{r}_{s}(M)  \, , \, \forall  r,s \in \mathbb{N}
  \, : \,(r,s ) \neq (0,0)
\end{equation}
\end{itemize}

\smallskip

 Since in the former situation the Gr\"{o}newald-Van Hove Theorem
\cite{Abraham-Marsden-78}, \cite{Guillemin-Sternberg-84} gives a
conceptually satisfying characterization of the involved ordering
problem stating the impossibility of making a consistent canonical
quantization of any polynomial in q and p, it appears then natural
to investigate whether an analogous theorem holds in the latter
situation, namely in General Relativity.

This is indeed the case as it may be appreciated as soon as one
performs a simple formal recasting of well-known arguments
concerning curvature terms related to the ordering problems.
\newpage
\section{The  Gr\"{o}enewald Van Hove theorem}
Let us consider the symplectic manifold $ ( T^{\star} \mathbb{R} ,
\omega ) $ where of course $  T^{\star} \mathbb{R} =
\mathbb{R}^{2}$ while $ \omega := d p \wedge d q $ is the
canonical symplectic form.

Let us denote by $ \mathcal{U} $ the Lie algebra of the
real-valued polynomials over $ \mathbb{R}^{2} $ where the Lie
brackets are the Poisson brackets associated to $ \omega $.

Given the Hilbert space $ \mathcal{H} := L^{2} ( \mathbb{R}, d
\mu_{Lebesgue}) $ let us introduce the following:
\begin{definition} \label{def:quantization}
\end{definition}
\emph{quantization:}

a map $ \hat{\cdot} : \mathcal{U} \mapsto \mathcal{L}_{s.a.} (
\mathcal{H} ) $ such that:
\begin{enumerate}
    \item for each finite subset $ S \subset \mathcal{U} $ there
    is a dense subspace $ \mathcal{D}_{S} \subset \mathcal{H} $
    such that:
\begin{equation*}
   \mathcal{D}_{S} \; \subset   \mathcal{D}_{f} \; and \; \hat{f}
   \mathcal{D}_{S} \subset  \mathcal{D}_{S} \; \; \forall f \in \mathcal{U}
\end{equation*}
    \item $ \widehat{ f + g }  \; = \; \hat{f} + \hat{g} $
    pointwise on $ \mathcal{D}_{S} \; \; \forall f \in S $
    \item $ \widehat{ \lambda f} \; = \; \lambda \hat{f} \; \; \forall \lambda \in \mathbb{R} $
    \item $ \widehat{ \{ f , g \} } \; = \; \frac{1}{i} [ \hat{f} ,
    \hat{g} ] $ on $ \mathcal{D}_{S} $
    \item $ \hat{I} \; = \; I $
     \item $ \hat{q} = $ multiplication by q and $ \hat{p} =
     \frac{1}{i} \frac{d}{d q} $
\end{enumerate}

\begin{theorem} \label{th:Groenwald Van Hove theorem}
\end{theorem}
GR\"{O}ENEWALD VAN-HOVE THEOREM

It doesn't exist a quantization map $ \hat{\cdot} $.

\smallskip

\begin{remark} \label{eq:quantization can't go over second order}
\end{remark}
Indeed the proof of theorem\ref{th:Groenwald Van Hove theorem},
(for which we demand to
\cite{Abraham-Marsden-78},\cite{Guillemin-Sternberg-84}) shows
that the quantization map $ \hat{\cdot} $ can be defined only for
the Lie subalgebra $ \mathcal{U}_{2} $ of the polynomials of
degree less or equal than 2 resulting in the operators:
\begin{equation}
  \widehat{q^{2}} \; = \; \hat{q}^{2}
\end{equation}
\begin{equation}
  \widehat{p^{2}} \; = \; \hat{p}^{2}
\end{equation}
\begin{equation}
  \widehat{q p} \; = \; \frac{1}{2} ( \hat{q} \hat{p} + \hat{p}
  \hat{q})
\end{equation}
but cannot be extended consistently to the  set  $ \mathcal{U}_{3}
$ of the polynomials of degree less or equal than 3 \footnote{Let
us observe that   $ \mathcal{U}_{3} $ is not a Lie subalgebra of
 $ \mathcal{U}_{2} $ since, for instance, $ \{ q^{3} , p^{3} \} \; = \; 9 q^{2} p^{2} \notin \mathcal{U}_{3}
 $}.

\smallskip

\begin{remark} \label{rem:Weyl ordering}
\end{remark}
One can of course define the \emph{Weyl ordering map} $
\hat{\cdot}^{W.O} : \mathcal{U} \mapsto \mathcal{L}_{s.a.} (
\mathcal{H} ) $ defined by symmetrizing the product of operators
in all possible combinations with equal weight:
\begin{equation}
 \widehat{ q^{2} p }^{W.O.} \; := \; \frac{1}{3} ( \widehat{q^{2}}
 \hat{p} + \hat{q} \hat{p} \hat{q} + \hat{p} \hat{q^{2}})
\end{equation}
and so on.

As it is well-known, given a classical hamiltonian H(q,p), this
corresponds to the \emph{mid-point-prescription}:
\begin{equation}
    < q' | \hat{H}^{W.O.} | q > \; = \; \int_{- \infty}^{+ \infty}
    \frac{d p}{2 \pi } \exp[ - i p ( q - q') ] H ( \frac{q+ q'
    }{2},p )
\end{equation}

The key point is that the map $ \hat{\cdot}^{W.O} $ is not a
\emph{quantization} (i.e. it doesn't respects all the conditions
required by the definition\ref{def:quantization}).

Alternatively  one can introduce the \emph{normal ordering map} $
\hat{\cdot}^{N.O} : \mathcal{U} \mapsto \mathcal{L}_{s.a.} (
\mathcal{H} ) $ defined by the condition that the $ \hat{p} $
operators are always put to the left of the $ \hat{q} $ operators,
i.e.:
\begin{equation}
    \widehat{q p}^{N.O.} \; := \; \hat{p} \hat{q}
\end{equation}
\begin{equation}
      \widehat{q^{2} p}^{N.O.} \; := \; \hat{p} \hat{q^{2}}
\end{equation}
and so on.

As it is well-known, given a classical hamiltonian H(q,p), this
corresponds to the \emph{right-point-prescription}:
\begin{equation}
    < q' | \hat{H}^{N.O.} | q > \; = \; \int_{- \infty}^{+ \infty}
    \frac{d p}{2 \pi } \exp[ - i p ( q - q') ] H ( q ,p )
\end{equation}

Let us observe that not only $ \hat{\cdot}^{N.O} $ is not a
quantization but even its restriction to $ \mathcal{U}_{2} $
doesn't obey the conditions required by the
definition\ref{def:quantization}.

\newpage
\section{The limits in the application of the Minimum Substitution Prescription}
Given the Minkowski space-time $ ( \mathbb{R}^{4} , \eta_{a b}) $
and a curved space-time $ ( M , g_{ab}) $ let us introduce the
following:
\begin{definition}
\end{definition}
\emph{generalization map}:

a map $ \overbrace{ \cdot} : \mathcal{T} ( \mathbb{R}^{4} )
\mapsto \mathcal{T} ( M) $:
\begin{enumerate}
    \item $ \overbrace{\mathcal{T}^{r}_{s} ( \mathbb{R}^{4} ) } \; \subseteq \;
    \mathcal{T}^{r}_{s} ( M ) \; \forall r,s \in\mathbb{N} $
    \item $ \overbrace{ X^{a_{1} \cdots a_{r}}_{b_{1} \cdots b_{s}} + Y^{a_{1} \cdots a_{r}}_{b_{1} \cdots b_{s}} } \; = \overbrace{ X^{a_{1} \cdots a_{r}}_{b_{1} \cdots b_{s}}} +
    \overbrace{Y^{a_{1} \cdots a_{r}}_{b_{1} \cdots b_{s}}} \; \; \forall X^{a_{1} \cdots a_{r}}_{b_{1} \cdots b_{s}} , Y^{a_{1} \cdots a_{r}}_{b_{1} \cdots b_{s}} \in \mathcal{T}^{r}_{s} (
    \mathbb{R}^{4}) \, , \, \forall r,s \in\mathbb{N} $
    \item $ \overbrace{ \lambda X^{a_{1} \cdots a_{r}}_{b_{1} \cdots b_{s}} } \; = \; \lambda
    \overbrace{X^{a_{1} \cdots a_{r}}_{b_{1} \cdots b_{s}}} \; \, \forall \lambda \in \mathbb{R} \, , \, \forall  X^{a_{1} \cdots a_{r}}_{b_{1} \cdots b_{s}} \in \mathcal{T}^{r}_{s} (
    \mathbb{R}^{4}) ,  \; \forall r,s \in\mathbb{N} $
    \item $ \overbrace{ X Y } \; = \; \overbrace{X}
    \overbrace{Y}  \; \; \forall X , Y \in
    \mathcal{T} (\mathbb{R}^{4}) $
    \item $ \overbrace{ \nabla^{(\eta)}_{a} X^{a_{1} \cdots a_{r}}_{b_{1} \cdots b_{s}}} \; = \; \nabla^{(g)}_{a} \overbrace{X^{a_{1} \cdots a_{r}}_{b_{1} \cdots b_{s}}} \;  \forall X^{a_{1} \cdots a_{r}}_{b_{1} \cdots b_{s}} \in \mathcal{T}^{r}_{s} (\mathbb{R}^{4}), \forall r,s \in \mathbb{N} $
    \item $ \overbrace{ \eta_{a b}  } \; = \; g_{a b}$
\end{enumerate}

Following the terminology of \cite{Wald-84} in  General Relativity
one would be tempted to assume that the generalization of a
non-gravitational law of Special Relativity, that is a
non-gravitational law holding in Minkowski-spacetime $ (
\mathbb{R}^{4} , \eta_{a b})$, to a non-gravitational law of
Physics holding on an arbitrary spacetime $ ( M , g_{a b} ) $ is
ruled by the following:
\begin{principle} \label{prin:Minimum Substitution Prescription}
\end{principle}
\emph{Minimum Substitution Prescription}
\begin{center}
\textbf{ if $   X^{a_{1} \cdots a_{r}}_{b_{1} \cdots b_{s}} = 0 $
is a non-gravitational Physics' law on $ ( \mathbb{R}^{4}, \eta_{a
b}) $ then $   \overbrace{ X^{a_{1} \cdots a_{r}}_{b_{1} \cdots
b_{s}}} = 0 $ is a Physics' law on $ ( M , g_{a b})$ where $
\overbrace{\cdot} $ is a suitable generalization map}.
\end{center}

\smallskip

\begin{example}
\end{example}
In the Minkowski spacetime $  ( \mathbb{R}^{4} , \eta_{a b})$
indices are raised and lowered by contraction with $ \eta_{a b} $
For instance:
\begin{equation}
    X_{a} \; = \; \eta_{a b} X^{b}
\end{equation}
Applying the Principle \ref{prin:Minimum Substitution
Prescription} it follows that on an arbitrary spacetime $ ( M ,
g_{a b} ) $ indices are raised and lowered by contraction with $
g_{a b} $. For instance:
\begin{equation}
    \overbrace{X_{a}} \; = \;\overbrace{ \eta_{a b} X^{b}} \; = \;
    g_{a b} \overbrace{ X^{b}}
\end{equation}

\smallskip

\begin{example}
\end{example}
The motion of a free particle of mass $m>0$  in the Minkowski
spacetime is given by:
\begin{equation}
    u^{a} \nabla^{(\eta)}_{a} u^{b} \; = \; 0
\end{equation}
where $ u^{a}$ is the \emph{4-velocity} of the particle (defined
as the tangent vector to its worldline parametrized through proper
time $ \tau $)  and hence the worldline of such a particle is a
time-like geodesic  of $ ( \mathbb{R}^{4}, \eta_{a b}) $.

Applying the Principle\ref{prin:Minimum Substitution Prescription}
it follows that the motion of a free-falling particle of mass $ m
>0 $ on an arbitrary curved spacetime $ ( M , g_{a b} ) $ is given
by:
\begin{equation}
   \overbrace{u^{a} \nabla^{(\eta)}_{a} u^{b}} \; = \; 0
\end{equation}
i.e.:
\begin{equation}
  \overbrace{u^{a}} \nabla^{(g)}_{a} \overbrace{u^{b}} \; = \; 0
\end{equation}
 It follows that the worldline of a massive free falling particle
 on an arbitrary curved spacetime $ ( M , g_{a b } ) $ is a
 time-like geodesic of $ ( M , g_{a b} ) $.

\smallskip

\begin{example}
\end{example}
A \emph{perfect fluid} on the Minkowski space-time $ (
\mathbb{R}^{4}, \eta_{a b}) $ is defined as a continuous
distribution of matter with stress-energy tensor $ T_{ab} $ of the
form:
\begin{equation}
    T_{a b} \; = \; \rho u_{a} u_{b} + P ( \eta_{a b} + u_{a}
    u_{b})
\end{equation}
where $ u^{a} $ is the unit time-like vector field representing
the 4-velocity of the fluid, and where $ \rho $ and $ P $ are,
respectively, the mass-energy density and the pressure of the
fluid measured in its rest frame.

In absence of external forces the equation of motion of a perfect
fluid in Minkowski space-time is:
\begin{equation}
    \nabla^{(\eta) a} T_{a b} \; = \; 0
\end{equation}
i.e.:
\begin{equation} \label{eq:equation of motion of a perfect fluid in Special Relativity}
   \nabla^{(\eta) a } [ \rho u_{a} u_{b} + P ( \eta_{a b} + u_{a}
    u_{b})] \; = \; 0
\end{equation}
Projecting eq.\ref{eq:equation of motion of a perfect fluid in
Special Relativity} parallel and perpendicular to $ u^{b} $ one
obtains that:
\begin{equation}
    u^{a} \nabla^{(\eta)}_{a} \rho + ( \rho + P) \nabla^{(\eta) a }
    u_{a} \; = \; 0
\end{equation}
\begin{equation}
    ( P + \rho ) u^{a} \nabla^{(\eta)}_{a} u_{b} + ( \eta_{a b} +
    u_{a} u_{b}  ) \nabla^{(\eta) a } P \; = \; 0
\end{equation}
Applying the Principle\ref{prin:Minimum Substitution Prescription}
it follows that a perfect fluid on a $ ( M , g_{a b} ) $ is
defined as a continuous distribution of matter with stress-energy
tensor:
\begin{equation}
    \overbrace{T_{a b}} \; = \; \overbrace{\rho u_{a} u_{b} + P ( \eta_{a b} + u_{a}
    u_{b})} \; = \; \overbrace{\rho} \overbrace{u_{a}}
    \overbrace{u_{b}} + \overbrace{P} ( g_{a b} + \overbrace{u_{a}}
\overbrace{u_{b}} )
\end{equation}
If a perfect-fluid is free-falling on  $( M , g_{a b}) $ its
equation of motion,  applying the Principle\ref{prin:Minimum
Substitution Prescription}, is:
\begin{equation}
    \overbrace{ \nabla^{(\eta) a } T_{a b}  } \; = \;
    \nabla^{(g) a } \overbrace{T_{a b}} \; = \; 0
\end{equation}
that leads to:
\begin{equation}
    \overbrace{u^{a} \nabla^{(\eta)}_{a} \rho + ( \rho + P )
    \nabla^{(\eta) a } u_{a}} \; = \; \overbrace{u^{a}}  \nabla^{(g)}_{a}
    \overbrace{\rho} + ( \overbrace{\rho }+ \overbrace{P} )
    \nabla^{(g) a } \overbrace{u_{a}} \; = \; 0
\end{equation}
\begin{equation}
    \overbrace{( P + \rho ) u^{a} \nabla^{(\eta)}_{a} u_{b} + ( \eta_{a b} +
    u_{a} u_{b}  ) \nabla^{(\eta) a } P }\; = \; ( \overbrace{P}
    + \overbrace{\rho}) \overbrace{u^{a}}  \nabla^{(g)}_{a}
    \overbrace{u_{b}} + ( g_{a b} + \overbrace{u_{a}}
    \overbrace{u_{b}}) \nabla^{(g) a} \overbrace{P} \; = \; 0
\end{equation}

\smallskip

\begin{example}
\end{example}
Electromagnetism on the Minkowski space-time  $ ( \mathbb{R}^{4},
\eta_{a b}) $ is described by the Maxwell equations:
\begin{equation} \label{eq:first Maxwell equation of Minkowski spacetime}
    \nabla^{(\eta) a } F_{a b} \; = \; - 4 \pi j_{b}
\end{equation}
\begin{equation}
   \nabla^{(\eta)}_{[a} F_{b c ]} \; = \; 0
\end{equation}
where the electromagnetic field $ F_{a b} $ is totally
antisymmetric:
\begin{equation} \label{totally antisymmetry of the electromagnetic field on Minkowski spacetime}
    F_{[a b]} \; =  \; F_{a b}
\end{equation}
Equation\ref{totally antisymmetry of the electromagnetic field on
Minkowski spacetime} implies that:
\begin{equation} \label{eq:conservation of the 4-current on an arbitrary spacetime}
   \nabla^{(\eta) b } j_{b} \; = \; - \frac{1}{4 \pi }  \nabla^{(\eta) b
   }  \nabla^{(\eta) a}  F_{a b} \; = \; 0
\end{equation}
so that the 4-current $ j^{b} $ is conserved.

 The stress-energy tensor of the electromagnetic
field is:
\begin{equation}
    T_{a b} \; = \; \frac{1}{4 \pi } ( F_{a c} F_{b}^{c} -
    \frac{1}{4} \eta_{a b } F_{d e} F^{d e} )
\end{equation}

The equation of motion of a particle of mass $ m > 0 $ and
electric charge q moving in the electromagnetic field $ F_{a b} $
is:
\begin{equation}
    u^{a} \nabla^{(\eta)}_{a} u^{b} \; = \; \frac{q}{m} F^{b}_{c} u^{c}
\end{equation}
Applying the Principle\ref{prin:Minimum Substitution Prescription}
it follows that Electromagnetism on a spacetime $ ( M , g_{ab} ) $
is described by the generalized Maxwell equations:
\begin{equation} \label{eq:first Maxwell equation on an arbitrary spacetime}
   \overbrace{ \nabla^{(\eta) a } F_{a b}  + 4 \pi j_{b}} \; = \;
   \nabla^{(g) a } \overbrace{F_{a b}} + 4 \pi \overbrace{j_{b}}
   \; = \; 0
\end{equation}
\begin{equation}
  \overbrace{ \nabla^{(\eta)}_{[a} F_{b c ]}} \; = \; \nabla^{(g)}_{[a} \overbrace{F_{b c ]}}  \; = \; 0
\end{equation}
where the electromagnetic tensor $ \overbrace{F_{ab}} $ is totally
antisymmetric:
\begin{equation} \label{totally antisymmetry of the electromagnetic field on a arbitrary spacetime}
   \overbrace{ F_{[a b]} - F_{a b}} \; = \;   \overbrace{F_{[a
   b]}} -\overbrace{ F_{a b}} \; = \; 0
\end{equation}
Let us observe that:
\begin{equation} \label{eq:conservation of the 4-current}
   \overbrace{\nabla^{(\eta) b } j_{b}} \; =  \; \nabla^{(g) b }
   \overbrace{j_{b}} \; = \;0
\end{equation}
so that the 4-current $\overbrace{ j_{b}}$ is conserved.

The stress-energy tensor of the electromagnetic field is:
\begin{equation}
   \overbrace{ T_{a b}} \; = \; \overbrace{\frac{1}{4 \pi } ( F_{a c} F_{b}^{c} -
    \frac{1}{4} \eta_{a b } F_{d e} F^{d e})} \; = \; \frac{1}{4
    \pi} ( \overbrace{F_{a c}} \overbrace{F_{b}^{c}} -
    \frac{1}{4} g_{a b } \overbrace{ F_{d e}} \overbrace{F^{d e}})
\end{equation}
The equation of motion of a particle of mass $ m > 0 $ and
electric charge q moving in the electromagnetic field $
\overbrace{F_{a b}} $ is:
\begin{equation}
    \overbrace{u^{a} \nabla^{(\eta)}_{a} u^{b} - \frac{q}{m} F^{b}_{c}
    u^{c}} \; = \; \overbrace{u^{a}} \nabla^{(g)}_{a} \overbrace{u^{b}} - \frac{q}{m} \overbrace{F^{b}_{c}}
    \overbrace{u^{c}} \; = \; 0
\end{equation}

\smallskip

That the Principle\ref{prin:Minimum Substitution Prescription}
cannot be assumed in its whole generality is, anyway, a
consequence of the following:

\begin{theorem} \label{th:Groenwald Van Hove like theorem of general relativity}
\end{theorem}
GR\"{O}ENWALD VAN HOVE LIKE THEOREM OF GENERAL RELATIVITY
\begin{center}
 it doesn't exist a \emph{generalization  map} $  \overbrace{ \cdot} $
\end{center}
\begin{proof}
Let us suppose, ad absurdum, that a \emph{generalization map}  $
\overbrace{\cdot}$ exists.

One has then that:
\begin{equation}
    \overbrace{ \nabla^{(\eta)}_{a} \nabla^{(\eta)}_{b} T^{c} } \; = \;
      \nabla^{g}_{a} \nabla^{g}_{b} \overbrace{T^{c}}  \; \; \forall T^{c}
    \in \mathcal{T}_{0}^{1} ( \mathbb{R}^{4} )
\end{equation}
From the other side:
\begin{equation}
     \overbrace{ \nabla^{(\eta)}_{a} \nabla^{(\eta)}_{b} T^{c} } \; =
     \; \overbrace{   \nabla^{(\eta)}_{b} \nabla^{(\eta)}_{a} T^{c}
     } \; = \;   \nabla^{g}_{b} \nabla^{g}_{a} \overbrace{T^{c}}  \; \; \forall T^{c}
    \in \mathcal{T}_{0}^{1} ( \mathbb{R}^{4} )
\end{equation}
Let us now observe that:
\begin{equation}
   \nabla^{(g)}_{a} \nabla^{(g)}_{b}\overbrace{ T^{c}} \; = \nabla^{(g)}_{b}
   \nabla^{(g)}_{a} \overbrace{T^{c}} \, - \,  R^{(g) c}_{a b d }\overbrace{ T^{d}} \; \; \forall T^{c}
    \in \mathcal{T}_{0}^{1} ( \mathbb{R}^{4} )
\end{equation}
and hence:
\begin{equation}
   \overbrace{ \nabla^{(\eta)}_{a} \nabla^{(\eta)}_{b} T^{c} }
    \; = \overbrace{ \nabla^{(\eta)}_{a} \nabla^{(\eta)}_{b} T^{c}
    } -   R^{(g) c}_{a b d } \overbrace{T^{d}} \; \; \forall T^{c}
    \in \mathcal{T}_{0}^{1} ( \mathbb{R}^{4} )
\end{equation}
that, being the Riemann curvature tensor $ R^{(g) c}_{a b d } \neq
0 $, is absurd.
\end{proof}

\smallskip

\begin{example}
\end{example}
Let us describe Electromagnetism over Minkowski spacetime $(
\mathbb{R}^{4} , \eta_{a b} ) $ in terms of the vector potential $
A^{a} $ defined by:
\begin{equation} \label{eq:definition of the vector potential in special relativity}
    F_{a b} \; =: \; \nabla^{(\eta)}_{a} A_{b} -  \nabla^{(\eta)}_{b} A_{a}
\end{equation}
Equation \ref{eq:first Maxwell equation of Minkowski spacetime}
becomes:
\begin{equation} \label{eq:first Maxwell equation expressed in terms of the vector potential in special relativity}
    \nabla^{(\eta) a} ( \nabla^{(\eta)}_{a} A_{b} -  \nabla^{(\eta)}_{b}
    A_{a} ) \; = \; - 4 \pi j_{b}
\end{equation}
The equation \ref{eq:definition of the vector potential in special
relativity} individuates $ A_{a} $ up to a gauge transformation:
\begin{equation}\label{eq:gauge transformation in special relativity}
    A_{a} \; \rightarrow \;   A_{a} + \nabla^{(\eta)}_{a} f \; \;
    f \in \mathcal{T}^{0}_{0} ( \mathbb{R}^{4} )
\end{equation}
By solving the equation:
\begin{equation}
  \nabla^{(\eta) a } \nabla^{(\eta)}_{a} f \; = -  \nabla^{(\eta) b
  } A_{b}
\end{equation}
we can make a gauge transformation to impose the \emph{Lorentz
gauge condition}:
\begin{equation}\label{eq:Lorentz gauge condition in special relativity}
  \nabla^{(\eta) a } A_{a} \; = \; 0
\end{equation}
In this gauge equation\ref{eq:first Maxwell equation expressed in
terms of the vector potential in special relativity} reduced to:
\begin{equation}
   \nabla^{(\eta) a } \nabla^{(\eta)}_{a} A_{b} \; = \; - 4 \pi j_{b}
\end{equation}
Applying the Principle\ref{prin:Minimum Substitution Prescription}
it follows that Electromagnetism on a spacetime $ ( M , g_{ab} ) $
may be described in terms of the generalized vector potential  $
\overbrace{A^{a}} $ defined by:
\begin{equation} \label{eq:definition of the vector potential in general relativity}
    \overbrace{F_{a b}} \; =: \; \overbrace{\nabla^{(\eta)}_{a} A_{b} -  \nabla^{(\eta)}_{b}
    A_{a}} \; = \;  \nabla^{(g)}_{a}\overbrace{ A_{b}} -  \nabla^{(g)}_{b} \overbrace{A_{a}}
\end{equation}
Equation\ref{eq:first Maxwell equation on an arbitrary spacetime}
becomes:
\begin{equation} \label{eq:first Maxwell equation expressed in terms of the vector potential in general relativity}
   \overbrace{ \nabla^{(\eta) a} ( \nabla^{(\eta)}_{a} A_{b} -  \nabla^{(\eta)}_{b}
    A_{a} ) + 4 \pi j_{b}} \; = \; \nabla^{(g) a} ( \nabla^{(g)}_{a} \overbrace{ A_{b}} -  \nabla^{(g)}_{b}
   \overbrace{ A_{a}} ) +  4 \pi \overbrace{ j_{b}} \; = \; 0
\end{equation}
Let us now observe that applying the  Principle\ref{prin:Minimum
Substitution Prescription} it follow that in the Lorentz gauge:
\begin{equation}
 \overbrace{ \nabla^{(\eta) a} A_{a} } \; = \; \nabla^{(g) a}
\overbrace{ A_{a} } \; = \; 0
\end{equation}
it should be:
\begin{equation}
   \overbrace{\nabla^{(\eta) a } \nabla^{(\eta)}_{a} A_{b} + 4 \pi
   j_{b}} \; = \; \nabla^{(g) a } \nabla^{(g)}_{a} \overbrace{A_{b}} +  4 \pi
   \overbrace{j_{b}} \; = \; 0
\end{equation}
and hence:
\begin{equation}
  \overbrace{j_{b}} \; = \; - \frac{1}{4 \pi } \nabla^{(g) a}  \nabla^{(g)
  }_{a} \overbrace{ A_{b}}
\end{equation}
But then:
\begin{equation}
  \nabla^{(g) b} \overbrace{j_{b}} \; = \; - \frac{1}{4 \pi } \nabla^{(g)
  b} \nabla^{(g) a}  \nabla^{(g)
  }_{a} \overbrace{ A_{b}} \; = - \frac{1}{4 \pi} \nabla^{(g) b} R^{(g) d}_{b}
  A_{d} \; \neq \; 0
\end{equation}
(where $ R^{(g)}_{a c} := R^{(g) b}_{a b c}  $ is the Ricci tensor
of the metric $ g_{d e} $) in contradiction with
equation\ref{eq:conservation of the 4-current on an arbitrary
spacetime}.

\smallskip

\begin{remark}
\end{remark}
We saw in the remark\ref{eq:quantization can't go over second
order} that a quantization map could be consistently defined on
the algebra $ \mathcal{U}_{2} $ of polynomials in q and p of order
less or equal than two, the inconsistencies stated by
Theorem\ref{th:Groenwald Van Hove theorem} arising as soon as one
tries to extend such a map to the set $ \mathcal{U}_{3} $ of
polynomials in q and p of order less or equal than 3.

In a similar way the proof of theorem\ref{th:Groenwald Van Hove
like theorem of general relativity} shows that defined the set of
tensors of order n over a differentiable manifold M as:
\begin{equation}
    \mathcal{T}_{n}(M) \; := \; \cup_{r,s \in \mathbb{N} : r+s \leq n}
    \mathcal{T}^{r}_{s}(M)
\end{equation}
a generalization map could be consistently defined on the set $
\mathcal{T}_{2}(\mathbb{R}^{4}) $ (it is essential, with this
regard, that $ \nabla^{(g)}_{a}  \nabla^{(g)}_{b} f =
\nabla^{(g)}_{b}  \nabla^{(g)}_{a} f \; \; \forall f \in
\mathcal{T}^{0}_{0}(M)$), the inconsistencies arising as soon as
one tries to extend such a map to the set $
\mathcal{T}_{3}(\mathbb{R}^{4}) $.

\begin{remark}
\end{remark}
One could think, with analogy to the Weyl-ordering of the remark
\ref{rem:Weyl ordering}, to define a map $ \overbrace{
\cdot}^{symmetric} : \mathcal{T} ( \mathbb{R}^{4} ) \mapsto
\mathcal{T} ( M) $  defined by symmetrizing the double product of
covariant derivatives  in all possible combinations with equal
weight, i.e.:
\begin{equation}
  \overbrace{ \nabla^{(\eta)}_{a}
  \nabla^{(\eta)}_{b} T^{c}}^{symmetric} \; := \; \frac{1}{2} (  \nabla^{(g)}_{a}
  \nabla^{(g)}_{b} \overbrace{T^{c}} +  \nabla^{(g)}_{b}
  \nabla^{(g)}_{a} \overbrace{T^{c}}) \; \; \forall T^{c}
    \in \mathcal{T}_{0}^{1} ( \mathbb{R}^{4} )
\end{equation}
But then:
\begin{equation}
    \overbrace{ \nabla^{(\eta)}_{a}
  \nabla^{(\eta)}_{b} T^{c}}^{symmetric} \; = \; \frac{1}{2} ( 2 \nabla^{(g)}_{b}
   \nabla^{(g)}_{a} \overbrace{T^{(c)}} \, - \,  R^{(g)c}_{ a b d }
   \overbrace{T^{d}} ) \; \; \forall T^{c}
    \in \mathcal{T}_{0}^{1} ( \mathbb{R}^{4} )
\end{equation}
and hence $ \overbrace{\cdot}^{symmetric} $ is not a
generalization map.
\newpage
\section{Acknowledgements}
I acknowledge funding related to a Marie Curie post-doc fellowship
of the EU network on "Quantum Probability and Applications in
Physics, Information Theory and Biology" contract
HPRNT-CT-2002-00279 (prolonged of two months). I would like to
thank prof. A. Khrennikov for stimulating discussions; of course
he has no responsibility of any error contained in these pages.
\newpage
\appendix
\section{Units and dimensions}
Following once more \cite{Wald-84} we adopt \emph{Planck units}
defined by the condition $ \hbar \; = \; G \; = c \; = 1 $. In
\emph{Planck units} all the quantities having International
System's dimensions expressible in terms of L , T and M are
dimensionless. In particular all lengths are expressed as
dimensionless multiples of the \emph{Planck length} $ l_{p} := (
\frac{ G \hbar }{ c^{3} })^{\frac{1}{2}} $.
\newpage
\section{Notation}
\begin{center}
  \begin{tabular} {|c|c|}
  $ \mathcal{L}_{s.a.} ( \mathcal{H} ) $ & self-adjoint operators
  over the Hilbert space $ \mathcal{H} $ \\
   $\nabla^{(g)}_{a}$ & Levi Civita covariant derivative of the metric $ g_{b c} $ \\
  $\mathcal{T} (M)$ & tensor fields over M \\
  $\mathcal{T}^{q}_{r} (M)$  & tensor fields of type (q,r) over M
  \\
  $ \mathcal{T}_{n}(M) $ & tensor fields of order n over M \\
  $ R^{(g) a}_{b c d} $ & Riemann curvature tensor of the metric $ g_{e f} $  \\
  $ R^{(g)}_{a b} $ & Ricci tensor of the metric $ g_{c d} $ \\
  $ T_{(a_{1} \cdots a_{n})} $ & totally symmetric tensor of type
  (0,n)  \\
 $ T_{[a_{1} \cdots a_{n}]} $ & totally antisymmetric tensor of type
  (0,n)  \\ \hline

\end{tabular}
\end{center}

\end{document}